\documentstyle[sprocl,epsfig]{article}

\def\Journal#1#2#3#4{{#1} {\bf #2}, #3 (#4)}

\def\NPB{{\em Nucl. Phys.} B }
\def\NPBPS{{\em Nucl. Phys.} B (Proc. Suppl.) }
\def\PLB{{\em Phys. Lett.}  B }
\def\PRL{{\em Phys. Rev. Lett.} }
\def\PRD{{\em Phys. Rev.} D }
\def\ZPC{{\em Z. Phys.} C }
\def\EPJ{{\em Eur. Phys. J.} C }

\def\ra{\rightarrow}

\def\al{\alpha}

\def\be{\begin{equation}}
\def\ee{\end{equation}}
\def\bea{\begin{eqnarray}}
\def\eea{\end{eqnarray}}
\def\ba{\begin{eqnarray*}}
\def\ea{\end{eqnarray*}}
\def\ra{\rightarrow}
\def\epm{e^+ e^-}
\def\pipi{\pi^+ \pi^-}
\def\amu{a_\mu}
\def\amuh{a_\mu^{{\mathrm had}}}
\def\MZ{M_Z}

\def\aiz{\alpha^{-1}(\MZ)}
\def\dalf{\Delta\alpha}
\def\das{\Delta\alpha(s)}

\def\dahs{\Delta\alpha^{(5)}_{\rm had}(s)}
\def\dahz{\Delta\alpha^{(5)}_{\rm had}(\MZ^2)}
\newcommand{\lsim}{\mbox{\raisebox{-0.3ex}{%
\footnotesize $\:\stackrel{<}{\sim}\:$}} }

\def\damu{\delta \amu}
\def\dalp{\delta \dalf}
\newcommand{\crn}{\nn \\}
\newcommand{\nn}{\nonumber}
\newcommand{\noi}{\noindent}

\newcommand{\gv}{\mbox{GeV}}

\newcommand{\MOM}{${\mathrm{MOM}}$ }

\newcommand{\MSb}{$\overline{\mathrm{MS}}$ }
\newcommand{\MSbm}{\overline{\mathrm{MS}} }
\newcommand{\veps}{\varepsilon}

\begin{document}
\parindent 0mm
\parskip 2mm
\renewcommand{\arraystretch}{1.4}

\vspace*{-20mm}

\thispagestyle{empty}
\hfill {\sc DESY 99-007} \qquad { } \par
\hfill January 1999 \qquad { }

\vspace*{10mm}

\title{HADRONIC EFFECTS IN $(g-2)_\mu$ and
$\alpha_{\small \mathrm{QED}}(M_Z)$:\\
STATUS AND PERSPECTIVES\footnote{Presented at the
{\em IVth International Symposium on
                   Radiative Corrections}, Barcelona, September 1998
                   (to appear in the proceedings, ed.~J. Sol\`a);}}

\author{F. JEGERLEHNER}

\address{Deutsches Elektronen Synchrotron DESY, Platanenallee 6,
D-15738 Zeuthen, Germany\\E-mail: fjeger@ifh.de}

\maketitle

\abstracts{
I review recent evaluations of the hadronic contributions to the muon
anomalous magnetic moment $(g-2)_\mu$ and to the effective fine
structure constant $\alpha(M_Z)$. A new estimate for the hadronic
shift $\dahz = 0.027782 \pm 0.000254$ is presented which implies
$\aiz=128.913 \pm 0.035$. It is based on a recent perturbative
calculation of the Adler function which includes mass effects up to
three-loops in a \MOM scheme and requires little ad hoc assumptions.
I then discuss perspectives for possible improvements in
estimations of $\amuh$ which we expect from the $\Phi$--factory {\sf
Daphne} at Frascati.}
\section{Status}
I briefly review the status of the estimations of hadronic {\em vacuum
polarization effects}. I have presented a similar report some time ago
in Ref.~\cite{Rhein} and the present contribution should be considered
as an Addendum to my previous summary. Vacuum polarization effects play
a role in many places in physics (charge screening). Their precise
knowledge is most important for the interpretation of high energy
precision experiments where theoretical predictions depend on the
effective fine structure constant $\alpha(s)=\alpha/(1-\das)$ as well
as for low energy precision experiments like for the anomalous
magnetic moment $\amu$, one of the most precisely measured quantities
in physics. The photon vacuum polarization amplitude $\Pi'_{\gamma}
(q^2)$ is defined by
\bea
\Pi^{\gamma}_{\mu \nu}(q) &=& i \int d^4 x e^{iqx} <0|T J^\gamma_\mu\;
(x)\; J^\gamma_\nu\;(0)\;|0> \crn &=& -\left(q^2\, g_{\mu \nu}- q_\mu
q_\nu\right)\; \Pi'_{\gamma}\;(q^2)
\label{CC}
\eea
and determines the full photon propagator
\bea
-g_{\mu\nu}\frac{i}{q^2} \frac{1}{1+e^2 \Pi'_\gamma(q^2)}
\eea
as well as the shift in the fine structure constant
\bea
\Delta
\alpha(q^2)=e^2\left(\Pi'_\gamma(0)-\Pi'_\gamma(q^2)\right)\;\;.
\eea
The latter is subtracted at zero momentum, i.e., in the classical
limit. At the one--loop order in perturbation theory one obtains
\bea
e^2 \Pi'_\gamma(q^2)=\frac{\alpha}{3\pi} \sum_f Q_f^2 N_{cf} \left(\ln
\frac{\mu^2}{m_f^2}+
\frac{5}{3}+y+\left(1+\frac{y}{2} \right)\:\beta
\ln \frac{\beta-1}{\beta+1} \right)
\eea
with $\mu$ the \MSb scale, $N_{cf}$ the color factor, $y=4m_f^2/s$ and
$\beta=\sqrt{1-y}$.

The electromagnetic current $J^\gamma_\mu$ is the sum of a leptonic
and a hadronic part. The leptonic part can be calculated in
perturbation theory and is known in QED up to
three--loops~\cite{MS98}: $\Delta \al_{\mathrm{lep}}=0.031497687$
(2--loop fraction $\sim 2.5 \times 10^{-3}$, 3--loop fraction $\sim
3.4 \times 10^{-5}$). The quark contribution cannot be calculated
reliably in perturbation theory because of low energy strong
interaction effects. Fortunately existing $\epm$ data allow us to
evaluate the hadronic contributions by means of the dispersion
integral
\bea
\dahs = - \frac{\alpha s}{3\pi}\;\bigg({\rm
P}\!\!\!\!\!\!  \int_{4m_\pi^2}^{E^2_{\rm cut}} ds'
\frac{R^{\mathrm{data}}_\gamma(s')}{s'(s'-s)} + {\rm P}\!\!\!\!\!\!
\int_{E^2_{\rm cut}}^\infty ds'
\frac{R^{\mathrm{pQCD}}_\gamma(s')}{s'(s'-s)} \bigg)
\label{DRalp}
\eea
where
\bea
R_\gamma(s) \equiv \frac{\sigma(e^+e^- \rightarrow \gamma^*
\rightarrow {\rm hadrons})}{ \sigma(e^+e^- \rightarrow \gamma^* \rightarrow
\mu^+ \mu^-)} = 12\pi{\rm Im}\Pi'_{\gamma}(s)
\eea
has been measured in $e^+e^-$--annihilation experiments up to about
$E_{\rm cut} \sim 40$ GeV. By virtue of the asymptotic freedom of QCD
the high energy tail can be safely calculated using pQCD~\cite{GKL}.

Taking a conservative attitude we have compiled the existing
$e^+e^-$--data (non--perturbative) and evaluated Eq.~(\ref{DRalp}) for
$\sqrt{s} \lsim E_{{\rm cut}}\sim 40 {\rm GeV}$. After including the
perturbative tail we found~\cite{EJ95}
\ba \dahz =
0.02804 \pm 0.00064~~~~~~~~ \mbox{(Eidelman, Jegerlehner 95)}
\ea
which leads to $\aiz=128.89 \pm 0.09$. It is obviously safe to use
pQCD in the continuum above the $\psi$ resonances and below the
$\Upsilon$ resonances (from 5.0 GeV to $M_\Upsilon$) and in the
continuum above the $\Upsilon$ resonances above about 12
GeV~\cite{ChK95}.  We then obtain~\cite{EJ95} $\dahz = 0.02801 \pm
0.00058$. Utilizing the fact that the vector--current hadronic
$\tau$--decay spectral functions are related to the iso--vector part of
the $\epm$--annihilation cross--section via an iso-spin
rotation~\cite{Tsai,EI95}, one can obtain information
from $\tau$--decay spectra to reduce the uncertainty of $\epm$
hadronic cross--sections at energies below the
$\tau$--mass~\cite{aleph,opal}. The inclusion of the $\tau$ data,
first performed in Ref.~\cite{ADH98}, unfortunately lead to a marginal
improvement only
\ba \dahz =
0.02810 \pm 0.00062~~~~~~~~ \mbox{(Alemany, Davier, H\"ocker 98)}
\ea
\begin{figure}
\centering
\begin{tabular}{|c|}
\hline
~~\\
\mbox{%
\epsfig{file=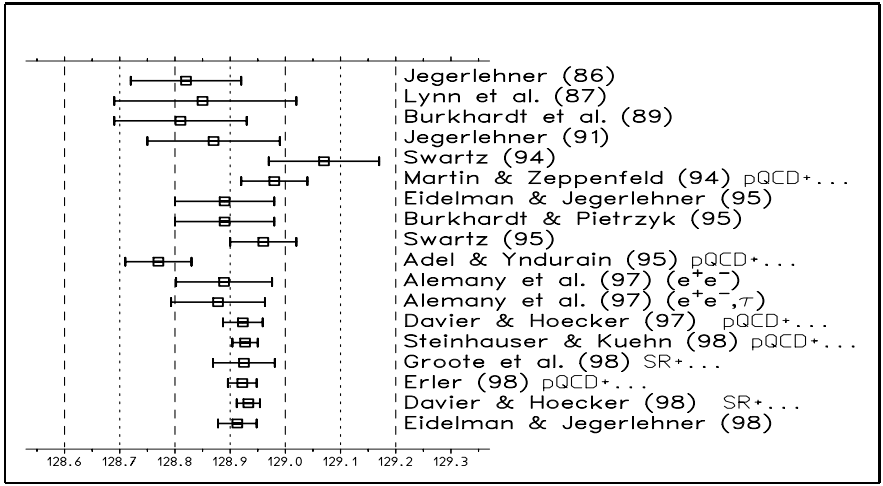 
        ,height=5.6cm  
        ,width=11.2cm   
       }%
}\\
\hline
\end{tabular}
\caption{Evaluations of $1/\alpha(M_Z)$.}
\label{fig:one}      
\end{figure}
The large uncertainty of about 2.3\% is due to large systematic errors
of the total hadronic cross--section measurements. To overcome this
disturbing fact more recent
estimations~\cite{MZ95,DH98a,KS98,GKSN98,DH98b} rely more on pQCD than
on data. Since hadronic $\tau$--decays seem to be described very well
by pQCD one assumes that perturbation theory describes the quantity
$R_\gamma(s)$ down to $M_\tau\sim 1.8$ GeV with good accuracy as well,
i.e., $ E_{{\rm cut}}\sim 1.8 {\rm \ GeV}$ is now a popular choice.
The low energy tail below 1.8 GeV, the narrow resonances ($\omega$,
$\phi$, $J/\psi$'s and $\Upsilon$'s) and
the charm resonance region from 3.7 to 5.0 are included separately,
still using experimental data. While the authors of Ref.~\cite{DH98a}
use the published charm data directly, obtaining~\cite{DH98a}
\ba \dahz =
0.02778 \pm 0.00026~~~~~~~~ \mbox{(Davier, H\"ocker 98a)}
\ea
the authors of Refs.~\cite{MZ95,KS98} utilize the charm data only
after re--normalizing individual data sets by factors accounting
for the mismatch in normalization:
$<R^{\mathrm{data}}_\gamma(s)/R^{\mathrm{pQCD}}_\gamma(s)>$ averaged
over small intervals just below and just above the charmonium
resonance domain. The result here is~\cite{KS98}
\ba \dahz =
0.02777 \pm 0.00017~~~~~~~~ \mbox{(K\"uhn, Steinhauser 98)}
\ea
Another approach is based on techniques which admit suppressing
contributions from data in problematic regions by a contour
trick~\cite{GKSN98,DH98b}. This yields~\cite{DH98b}
\ba \dahz =
0.02763 \pm 0.00016~~~~~~~~ \mbox{(Davier, H\"ocker 98b)}
\ea
and $\aiz=128.933 \pm 0.021$.

In most cases more or less obvious theoretical uncertainties typical
in hadron physics are not included because they are hard to be
specified. The recent reevaluations are collected together
with older results in Fig.~1.\\
\begin{figure}
\centering
\begin{tabular}{|c|}
\hline
~~\\
\mbox{%
\epsfig{file=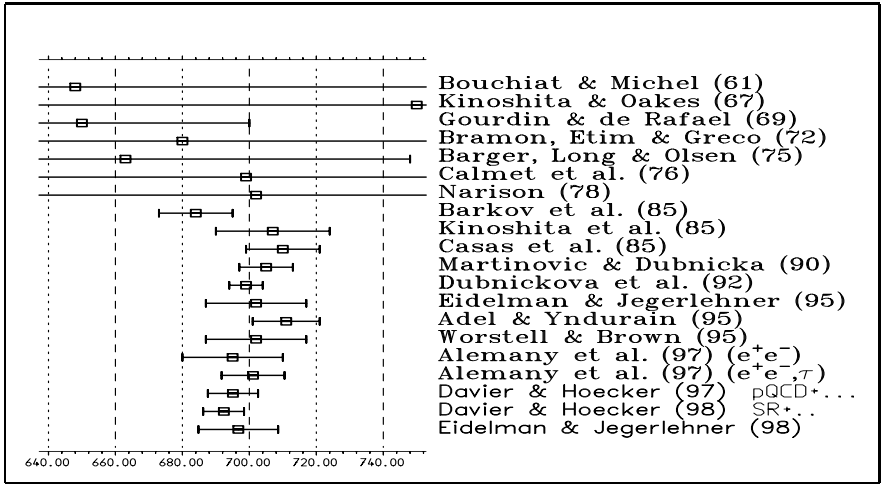 
        ,height=5.6cm  
        ,width=11.2cm   
       }%
}\\
\hline
\end{tabular}
\caption{Different estimates of $\amu^{\mathrm{had}}\times 10^{10}$.}
\label{fig:two}      
\end{figure}
The second quantity of interest is the hadronic contribution to
$\amu \equiv \frac{g_\mu -2}{2}$
which is determined from the same data by the dispersion integral
\bea
\amu = \left(\frac{\alpha m_\mu}{3\pi}
\right)^2 \bigg(\,\int\limits_{4 m_\pi^2}^{E^2_{\rm cut}}ds\,
\frac{R^{\mathrm{data}}_\gamma(s)\;\hat{K}(s)}{s^2}+
\int\limits_{E^2_{\rm cut}}^{\infty}ds\,
\frac{R^{\mathrm{pQCD}}_\gamma(s)\;\hat{K}(s)}{s^2}\, \bigg)
\label{DRamu}
\eea
The kernel $\hat{K}(s)$ is a known smooth bounded function growing from
0.63... at threshold to 1 at $\infty$. Note the extra $1/s$--enhancement
of contributions from low energies in Eq.~(\ref{DRamu})
as compared to Eq.~(\ref{DRalp}). Our updated evaluation reads
\ba \amuh =
(696.7 \pm 11.9) \times 10^{-10}~~~~~~~~ \mbox{(Eidelman, Jegerlehner 98)}
\ea
For this observable the use of $\tau$ spectral--functions in addition
to $\epm$ cross--sections leads to a dramatic reduction of the
uncertainty~\cite{ADH98}. Extended use of pQCD~\cite{DH98a} and application of
QCD sum rule techniques~\cite{DH98b} leads to further improvements
\ba \amuh &=&
(701.1 \pm 9.4) \times 10^{-10}~~~~~~~~ \mbox{(Alemany, Davier,
H\"ocker 98)}\\
\amuh &=&
(695.1 \pm 7.5) \times 10^{-10}~~~~~~~~ \mbox{(Davier, H\"ocker 98a)} \\
\amuh &=&
(692.4 \pm 6.2) \times 10^{-10}~~~~~~~~ \mbox{(Davier, H\"ocker 98b)}
\ea
Again recent reevaluations are shown together with older results in Fig.~2.
For a theory update of properties of $\amu$ we refer to Ref.~\cite{CM98}.
\section{Comments on the improvements}
{\bf 1)} A substantial reduction of uncertainties of $\amuh$ is possible by
using $\tau$--decay spectra to reduce the uncertainty of $\epm$
hadronic cross--sections at energies below the
$\tau$--mass~\cite{aleph,opal}. Conservation of the
iso--vector current (CVC) allows us to relate non--strange
$\tau$--decay data to $\epm$--annihilation data by an iso-spin
rotation:
$$ \tau^- \ra X^- \nu_\tau\;\;\; \leftrightarrow \;\;\; e^+ e¯ \ra X^0$$
where $X^-$ and $X^0$ are related hadronic states. The $\epm$
cross--section is then given by
\bea
\sigma_{\epm \ra X^0}^{I=1}= \frac{4 \pi \al^2}{s}v_{1,X^-}\;\;,\;\;\;
\sqrt{s} \leq M_\tau
\eea
The $\tau$ spectral function $v_1$ is obtained from the normalized invariant
mass-squared distribution $\rho\:(X^-)\equiv (1/N_{X^-}\,dN_{X^-}/ds)$
of the $\tau$ vector channel $X^-\nu_\tau$ by
\bea
v_{1,X^-}=A\,
\frac{B(\tau^- \ra X^-\nu_\tau)}{B(\tau^- \ra e^-\nu_\tau\bar{\nu}_e)}\,
\rho\:(X^-)\,\bigg[\left(1-\frac{s}{M^2_\tau} \right)^2
\left( 1+\frac{2s}{M^2_\tau}\right) \bigg]^{-1}
\eea
where
\ba
A=\frac{M^2_\tau}{6|V_{ud}|^2\,(1+\delta_{\rm EW})}
\ea
\noi
with $|V_{ud}|=0.9752 \pm 0.0007$ the CKM mixing matrix element and
$\delta=0.0194$ the electroweak radiative corrections~\cite{MS86}.  With
the precision of the validity of CVC, this allows to improve the $I=1$
part of the $\epm$ cross--section which by itself is not a directly
measurable quantity. It mainly improves the knowledge of the $\pipi$
channel ($\rho$--resonance contribution) which is dominating in
$\amuh$ (72\%). Problems of combining data are not too serious for
this particular observable/mode
which is pure $I=1$ and hence does not require to separate data into
iso-spin components. The precision is then only a matter
of how well the corresponding exclusive channels can be
measured. With the error estimates discussed in ~\cite{ADH98} it was
possible to reduced the error of $\amu$ considerably.

In any case the following points should be kept in mind:\\
{\bf (i)} the usual problems of matching an averaging data has to be dealt
with. For details we refer to~\cite{EJ95,ADH98}.\\
{\bf (ii)} In general the use of $\tau$ data requires a
splitting of the $\epm$ data into the $I=0$ and $I=1$ parts.
There is no precise method known how
to do this, except for counting even and odd numbers of pions. In any
case this introduces an additional systematic error which is hard to
estimate in a clean manner.\\
{\bf (iii)}The previous point as well as the CVC assumption rely on
estimates of the size of iso-spin breaking effects. This problem is
absent when using $\epm$ data solely. Iso-spin breaking effects
originate from the difference $m_u-m_d$ in the masses of the $u$ and $d$
quarks and from electromagnetic interaction, i.e., an imperfect
treatment of QED corrections of the various exclusive channels,
electromagnetic contributions to the hadron masses and widths  etc.\\
In a recent paper~\cite{EI95} Eidelman and Ivanchenko estimate the
validity of CVC by a detailed comparison of $\tau$ channels,
$$\tau^- \ra X^-\nu_\tau\;,\;\;X^-=\pi^-\pi^0, \pi^-3\pi^0,
(3\pi)^-\pi^0,\cdots $$ with corresponding $\epm$ channels. The
results are given in the Table~1.\\

\newpage

\begin{center}
Table 1: Branching Ratios of $\tau^- \ra X^- \nu_{\tau}$ in percent.\\[2mm]
\begin{tabular}{|c|c|c|c|}
\hline
Hadronic  &    World Average &       CVC      \\
State X   &        1996      &   Predictions  \\
\hline

$\pi^- \pi^0$   & $25.24 \pm 0.16$ & $24.74 \pm 0.79$ \\
\hline
$\pi^- 3\pi^0$  &     $~1.14 \pm 0.14$   &  $~1.07 \pm 0.10$  \\
\hline
$(3\pi)^-\pi^0$ &  $~4.25 \pm 0.09$  &   $~4.36 \pm 0.55$ \\
\hline
Total        & $30.63 \pm 0.23$ &    $30.17 \pm 1.00$  \\
\hline
\end{tabular}
\end{center}

\vspace*{3mm}
\noi
The difference between the CVC prediction and
the measured branching ratio of $\tau$ to 2-$\pi$ and 4-$\pi$ states is
consistent with zero within the errors (-0.46 $\pm$ 1.00)\% or -1.5\%
relatively. There is no reason {\bf not} to take this 1.5\% into account
while calculating $R^{I=1}(s)$ from the $\tau$ spectra.\\

\noi
{\bf 2)} Substantial progress in pQCD calculations, which now includes
quark mass effects up to three--loops~\cite{mqcd3}, allows us to apply
pQCD in regions where mass effects are important. Many authors now assume
pQCD to be valid down to $M_\tau$, and hence that
$$\sigma_{\mathrm{tot}}(\epm \ra  {\mathrm{hadrons}}) \sim
\sigma_{\mathrm{tot}}(\epm \ra  {\mathrm{quarks \ and \ gluons}})$$
The assumption seems to be supported by
\begin{itemize}
\item the apparent applicability of pQCD to $\tau$ physics. In fact the
running of $\alpha_s(M_\tau) \ra \alpha_s(M_Z)$ from the $\tau$ mass
up to LEP energies agrees well with the LEP value. The estimated
uncertainty may be debated, however.
\item non--perturbative (NP) effects if parametrized as
prescribed by the operator product expansion (OPE) of
the electromagnetic current correlator~\cite{SVZ} are
seemingly small~\cite{FJ86,DH98a}.
\end{itemize}

\noi
{\bf 3)} Let us consider possible NP effects separately. A way to
parametrize NP effects at sufficiently large energies and away from
resonances is provided by the OPE applied to Eq.~(\ref{CC}).
Due to non--vanishing gluon and light
quark condensates~\cite{SVZ} one finds the leading power corrections
\bea
\label{NP}
\Pi_\gamma^{'\mathrm{NP}}(Q^2) &=& \frac{4\pi\al}{3} \sum\limits_{q=u,d,s}
Q_q^2 N_{cq}\,
\cdot \bigg[\frac{1}{12}
\left(1-\frac{11}{18}a\right)
\frac{<\frac{\alpha_s}{\pi} G G>}{Q^4} \crn
&+&2\,\left(1+\frac{a}{3} +\left(\frac{11}{2}-\frac34
l_{q\mu}
\right)\,a^2\right)
\frac{<m_q \bar{q}q>}{Q^4} \\
&+& \left(\frac{4}{27}a
+\left(\frac{4}{3}\zeta_3-\frac{257}{486}-\frac13 l_{q\mu}
\right)\,
a^2\right)
\sum\limits_{q'=u,d,s} \frac{<m_{q'} \bar{{q'}}{q'}>}{Q^4}\,\bigg] \crn
&+& \cdots
 \nn
\eea
where $a\equiv \alpha_s(\mu^2)/\pi$
and $l_{q\mu}\equiv\ln(Q^2/\mu^2)$. $<\frac{\alpha_s}{\pi} G G>$ and
$<m_q \bar{q}q>$ are the scale-invariantly defined condensates. The
terms beyond leading order in $\alpha_s$ have been calculated from the
results which were obtained in Refs.~\cite{NPho},~\cite{ChGS}
and~\cite{SCh}.  Sum rule estimates of the condensates
yield typically $<\frac{\alpha_s}{\pi} G G>\sim (0.389\; \gv)^4\;\:,\;\;
<m_q \bar{q}q >\sim -(0.098\; \gv )^4\;\;{\mathrm{for}}\;\;q=u,d
\;\;{\mathrm{and}}\;\;[<m_q \bar{q}q >\sim-(0.218\; \gv )^4
\;\;{\mathrm{for}}\;\;q=s\;]$ with rather large uncertainties (see
Fig.~4 below).

Note that the expansion Eq.~(\ref{NP}) is just a parametrization of the
high energy tail of NP effects associated with the existence of
non--vanishing condensates. There are other kind of NP
phenomena like bound states,
resonances, instantons and what else. The dilemma with Eq.~(\ref{NP}) is
that it works only for $Q^2$ large enough and it has been
successfully applied in heavy quark physics. It fails do describe NP
physics at lower $Q^2$, once it starts to be
numerically relevant pQCD starts to fail because of the growth of the
strong coupling constant (see Fig.~4 below).\\

\noi
{\bf 4)} Some improvements are based on assuming \underline{global duality}
 and using \underline{sum rules} (SR). Starting point is
the OPE
$$<h|J_\mu (x)\, J_\nu (0)|h'>\simeq \sum_i C_{\mu\nu\:i}(x)\,H_i$$
where $H_i=<h|{\cal O}_i (0)|h'>$ is a matrix element of an operator
${\cal O}_i$ between hadronic states $h, h'$, and $ C_{\mu\nu\:i}(x)$ are
the Wilson coefficients. The global duality assumption asserts that all
non--perturbative physics resides in the $H_i$'s and $
C_{\mu\nu\:i}(x)$ is given by pQCD. Phenomenological tests infer that
this works at the 10\% level. Much better tests are needed to
allow for precise confirmation as required for application in
precision calculations.\\
\noi
I would like to point out that the following schemes have {\bf no}
justification:
\begin{itemize}
\item
$\sigma_{\mathrm{had}}=\sigma_{\mathrm{pQCD}}+\sigma_{\mathrm{resonances}}$
\item local duality: i.e., duality applied to energy intervals
(resonance regions)\\
\end{itemize}
\noi
Another warning I would like to make concerns pitfalls in the use of
dispersion relations. Often one encounters arguments of the following type:
consider a function, $\Delta \Pi'(s)=\Pi'(s)-\Pi'(0)$ say, which is an
analytic function order by order in perturbation theory. Analyticity
then infers that the contour integral along a path shown in Fig.~3 vanishes.
\begin{figure}
\centering
\mbox{%
\epsfig{file=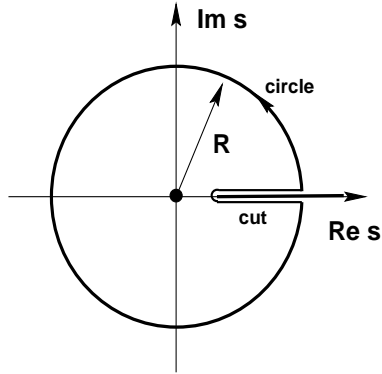 
        ,height=5cm  
        ,width=5cm   
       }%
}
\caption{Integral contour.} 
\label{fig:three}      
\end{figure}
Therefore
$$\int_{\mathrm{cut}} \frac{ds}{s}\Delta \Pi'(s)=-
\int_{\mathrm{circle}} \frac{ds}{s}\Delta \Pi'(s)$$ For a large enough
circle we can apply pQCD on the right hand side and thus obtain the
integral of our interest, which exhibits the non-perturbative
physics. What is usually forgotten is that the uncertainty is of the
order of $\delta=2\pi R \veps $  with $\veps$ being the small error
expected from the truncation of the perturbative series. $\delta$
easily can turn out to be large (due to $R$ large)
such that we are not able to make a safe estimate for the wanted
integral. Since analyticity
is true order by order in perturbation theory, we precisely reproduce
the perturbative answer for the left hand side if we use perturbation
theory on the right hand side. The additional use of the OPE to
include the condensates, in my opinion, does not make the estimate much
more convincing. While analyticity is a very powerful theoretical
concept it is difficult to be applied in numerical problems, because,
small perturbations in one place typically cause large variations at
remoter locations.
\section{My estimate of $\Delta\alpha^{(5)}_{\rm had}(-M_Z^2)$}
\begin{figure}[h]
\centering
\mbox{%
\epsfig{file=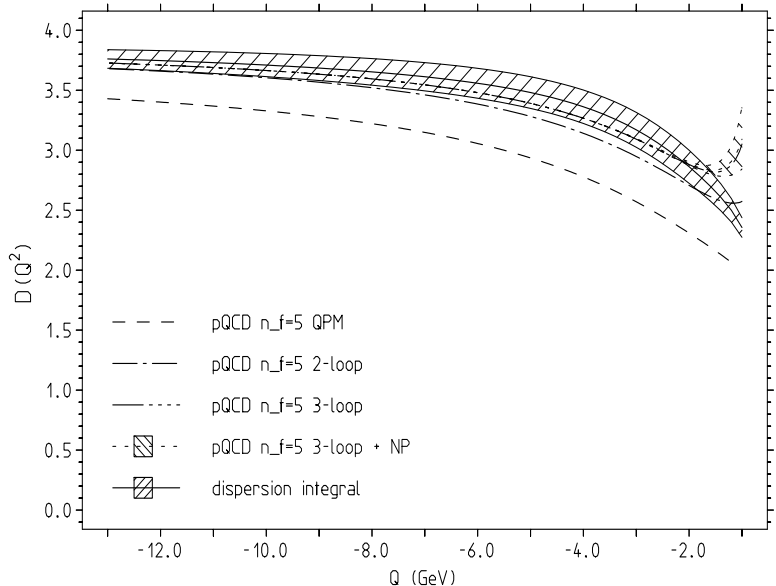 
        ,height=8cm  
        ,width=10.4cm   
       }%
}
\caption{Adler function: data versus theory (see Ref.$^{28}$).}
\label{fig:four}      
\end{figure}
Here I propose a method for estimating $\dahz$ which is free of
assumptions which are not under control. In a recent
paper~\cite{EJKV98} we investigated the validity of pQCD by means of
the Adler function in the Euclidean region where pQCD is supposed to
work best (far from resonances and thresholds).  Results obtained from
the evaluation of the dispersion integral are very well described by
pQCD down to about 2.5 GeV. In this analysis it turned out to be
crucial to take into account the exact mass dependence up to three
loops. For the three--loop case this was obtained by using the series
expansions presented in~\cite{mqcd3} together with Pad\'e improvement,
which from a practical point of view yields the exact behavior in the
Euclidean region. Because the precise mass--dependence is important we
utilize the gauge invariant background field \MOM scheme calculation
presented recently in Ref.~\cite{JT98}. The result is shown in Fig.~4.
With this result at hand, we then calculate in the Euclidean region
\be \Delta\alpha^{(5)}_{\rm had}(-M_Z^2)
=\left[\Delta\alpha^{(5)}_{\rm had}(-M_Z^2) -\Delta\alpha^{(5)}_{\rm
had}(-s_0)\right]^{\mathrm{pQCD}}+ \Delta\alpha^{(5)}_{\rm
had}(-s_0)^{\mathrm{data}}
\ee
and obtain, for $s_0=(2.5\, \gv)^2$
where $\Delta\alpha^{(5)}_{\rm had}(-s_0)^{\mathrm{data}} =0.007541 \pm
0.000254$,
\ba
\Delta\alpha^{(5)}_{\rm
had}(-M_Z^2) = 0.027737 \pm 0.000254\,{\mathrm{ \ and \ }}\,
\alpha^{-1}(-\MZ)=128.917 \pm 0.035
\ea
for the Euclidean effective fine structure constant a value
I would use in phenomenological applications without hesitation.
The virtues of this analysis are obvious:
\begin{itemize}
\item no problems with the physical threshold and resonances
\item pQCD is used only in the Euclidean region and not below 2.5 GeV.
 For lower scales pQCD ceases to describe properly the functional
 dependence of the Adler function~\cite{EJKV98} (although the pQCD answer
 remains within error bands down to about 1.6 GeV).
\item no manipulation of data must be applied and we need not use
 global or local duality. That contributions of the type Eq.~(\ref{NP})
 are negligible has been shown long time ago in Ref.~\cite{FJ86}. More
 recently this was confirmed in~\cite{DH98a}. This, however, does not
 proof the absence of other kind of non-perturbative effects.
\end{itemize}
Remaining problems are the following:\\ \noi
{\bf a)} contributions to the Adler function up to three--loops all have
the same sign and are substantial. Four-- and higher--orders could still
add up to non-negligible contribution. An error for missing higher
order terms is not included. The scheme dependence \MSb versus background
field \MOM has been discussed in Ref.~\cite{JT98}.\\ \noi
{\bf b)} the link between space--like and time-like region is the difference
$$\Delta=\dahz-\Delta\alpha^{(5)}_{\rm had}(-M_Z^2)
=0.000045 \pm 0.000002\,,$$ which can be
calculated in pQCD. It accounts for the $i\pi$--terms from the logs
$\ln (-q^2/\mu^2)=\ln(|q^2/\mu^2|)+i\pi$.  One may ask the question
whether these terms should be resummed at all, i.e., included in the
running coupling.  Usually such terms tend to cancel against constant
rational terms which are not included in the renormalization group
(RG) evolution. It is worthwhile to stress here that the running
coupling is {\bf not} a true function of $q^2$ (or even an analytic
function of $q^2$) but a function of the RG scale $\mu^2$. The
coupling as it appears in the Lagrangian in any case must be a
constant, albeit a $\mu^2$--dependent one, if we do not want to end up
in conflict with basic principles of quantum field theory.  The
effective identification of $\mu^2$ with a particular value of $q^2$
must be understood as a subtraction (reference) point.\\
The above result was obtained using the gauge invariant
background--field \MOM renormalization scheme, presented recently in
Ref.~\cite{JT98}. In the transition from the \MSb to the \MOM scheme
we adapt the rescaling procedure described in~\cite{JT98}, such that
for large $\mu$
\bea
\overline{\alpha}_s((x_0\mu)^2)=\alpha_s(\mu^2)+0+ O(\alpha_s^3)\;.
\eea
This means that $x_0$ is chosen such that the couplings coincide
to leading and next--to--leading order at asymptotically large
scales. Numerically we find $x_0\simeq 2.0144$. Due to this
normalization by rescaling the coefficients of the Adler--function
remain the same in both schemes up to three--loops. In the \MOM scheme
we automatically have the correct mass dependence of full QCD, i.e.,
we have automatic decoupling and do not need decoupling by hand and matching
conditions like in the \MSb scheme. For the numerical
evaluation we use the pole quark masses~\cite{PDG}
$m_c=1.55 \gv ,\;m_b=4.70 \gv,
\;m_t=173.80 \gv \;$ and the strong interaction coupling
$\alpha_{s\; {\small \MSbm}}^{(5)}(M_Z) = 0.120 \pm 0.003$.
For further details we refer to~\cite{EJKV98}.
\section{Improvements expected from Daphne}
For the rather dramatic progress on the theory--driven reduction of
uncertainties of $\dahz$ and $\amu$, discussed in Sec.~1, little
compelling theoretical or experimental justification is available to
remove doubts concerning the existence of further unaccounted
theoretical uncertainties. Therefore it is important to pursue the
experimental program to reduce uncertainties of $R_\gamma(s)$
measurements. New low energy data are expected from BES, CMD II and
DA$\Phi$NE in the near future. On a time scale of a few years the
$\Phi$--factory {\sf Daphne} is the most promising facility.
Therefore one of the challenges for {\sf Daphne} is presented by the
measurement of the total $e^+e^-$--hadronic cross--section at
$\sqrt{s}\lsim 2$ GeV, which provides the information on the hadronic
contributions to $(g-2)_\mu$ and to a lesser extent to
$\alpha(M_Z)$. Considerations from precision physics suggest that the
goal should be to measure the cross-section at the level of a few
permille. This could allow a reduction of the error on $\amu$ such as
to be closer to the contribution from virtual light-by-light
scattering (see e.g.~\cite{CM98}). To extract the total cross-section
a certain amount of modeling is necessary, since, in contrast to LEP,
at low energies the theoretical shape of the total cross-section
cannot be calculated from first principles. Important cross-checks
will be obtained by studying connections between form-factor
measurements at {\sf Daphne} and $\tau$-lepton decays (CLEO,
ALEPH, OPAL, DELPHI, BABAR) through iso-spin relations between
multi-pion final states, which we mentioned before.\\
\begin{figure}
\centering
\begin{tabular}{|c|}
\hline
~~\\
\mbox{%
\epsfig{file=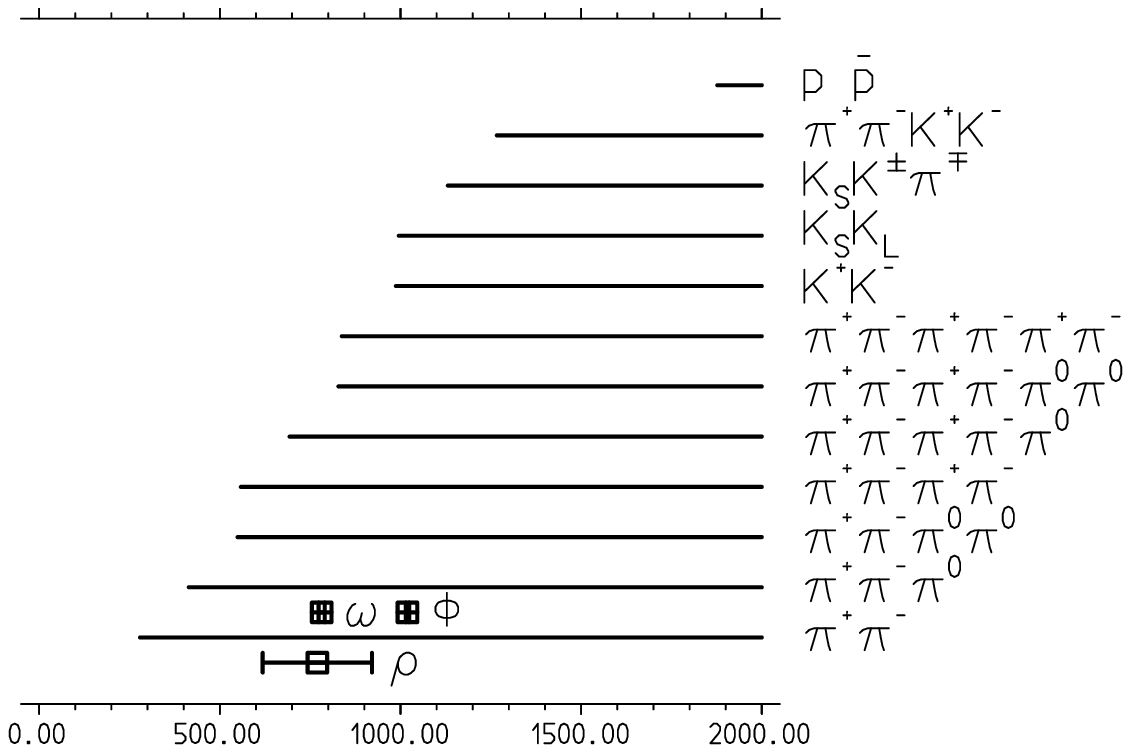 
        ,height=4.7cm  
        ,width=9.4cm   
       }%
}\\
\hline
\end{tabular}
\caption{Multi--particle channels in $e^+e^-$--annihilation at low energy.} 
\label{fig:five}      
\end{figure}
\noi
There are two stages: at the $\Phi$--resonance one can measure the
pion form factor $|F_\pi(s)|$ for $\sqrt{s}\lsim 1$ GeV via hard
photon tagging $$\Phi \ra \gamma^* + \gamma ({\mathrm{real, \
hard}})\ra \pipi + \gamma ({\mathrm{real, \ hard}})$$ At a second
stage, after a non--trivial {\sf Daphne} upgrade, one expects to be
able to perform an energy scan from threshold up to about 2 GeV, which
allows for a ``direct'' measurement of $R_\gamma(s)$. Unfortunately
the measurement of this inclusive quantity at low energies is not as
easy as at LEP/SLD. Because of competing 2 body channels
$e^+e^-,\;\mu^+\mu^-,\;\pi^+\pi^-,\;K^+K^-,\;\cdots\;,$
which must
be unambiguously separated, a good particle identification
is required. This means that the inclusive cross--section can be
obtained only by summing up the exclusive channels. This also requires
a detailed theoretical understanding of the individual channels at low
energies ($\lsim 2$ GeV). In Fig.~5 we show the spectrum of channels
which contribute the total hadronic cross--section. The
contributions from the exclusive channels to $\amuh$ are listed in
Table~2. In order to reduce the uncertainty from the region below 2
GeV to the 0.3\% level the accuracy required for the individual
channels is given in the last column.\\

\newpage

\begin{center}
Table~2: Contributions to $\tilde{a}_\mu^{\rm had}=\amuh\times 10^{10}$ from exclusive channels.\\
\begin{tabular}{|c|r|c||c|r|c|}
\hline
channel & $\tilde{a}_\mu^{\rm had}$ & acc. & channel & $\tilde{a}_\mu^{\rm had}$ & acc. \\
\hline
$ \rho, \omega \ra \pi^+ \pi^-$ & 506 & 0.3\% & $ 3\pi $ & 4& 10\% \\
$\omega \ra 3 \pi $ & 47 & $\sim$ 1\% &$K^+ K^-  $ & 4& $\downarrow$ \\
$\phi  $ &40 & $\downarrow$ & $K_S K_L  $ & 1& $\cdot$\\
$\pi^+ \pi^- \pi^0 \pi^0  $ & 24& $\cdot$ & $\pi^+ \pi^- \pi^+ \pi^- \pi^0$
& 1.8& $\cdot$\\
$\pi^+ \pi^- \pi^+ \pi^-   $ & 14& $\cdot$ & $\pi^+ \pi^- \pi^+ \pi^- \pi^+
\pi^-  $ & 0.5& $\cdot$\\
$\pi^+ \pi^- \pi^+ \pi^- \pi^0 \pi^0   $ & 5& 10\% &$p\bar{p}  $ & 0.2
& $\cdot$ \\
\hline
$ 2 {\rm \ GeV \ } \leq E \leq M_{J/\psi}  $ & 22 & &&&\\
$ M_{J/\psi} \leq E \leq M_{\Upsilon}  $ &20 & &&&\\
$M_{\Upsilon}< E  $ & $\lsim 5$  & &&&\\
\hline
\end{tabular}
\end{center}

\vspace*{5mm}
\noi
Theoretical work with the aim to calculate radiative corrections
at the level of precision as indicated in Table~2 is in progress.
\section{Summary and conclusions}
The future of precision physics depends to a large extent on the
uncertainties of hadronic vacuum polarization effects.
Present and future are different for $\amuh$ and $\dahz$:\\
\underline{$\amuh \,:$}~~~~uncertainty from $\pipi, \rho^0$ dominates\\[-2mm]
\begin{center}
Table~3: Uncertainties of $\amuh$ in units $10^{-11}$.\\[1mm]
\begin{tabular}{|c|l||c|l|}
\hline
$\damu$  & ~~~~~~input &
$\damu$  & ~~~~~~input \\
\hline
$\sim 156 $&$\epm$ data~\cite{EJ95,ADH98} &
$\sim 102 $&$\epm,\; \tau$  data~\cite{ADH98} \\
$\sim ~60 $& {\sf Daphne} + additional $\tau$ data &
$\sim ~40 $& BNL $\amu$--experiment~\cite{CM98} \\
$\sim ~62 $& theory-driven guess~\cite{DH98b} &&\\
\hline
\end{tabular}
\end{center}
\vspace*{3mm}
\noi
\underline{$\dahz \,:$}~~~~uncertainty distributed everywhere below
$\Upsilon$\\[2mm]
Uncertainties of $\epm$ data range from 5\% to 20\% at present which
yields $\dalp \sim 0.00065$ (data)~\cite{EJ95,ADH98}.
Suppose we could improve this to a 1\% measurement in systematics and
with high enough statistics, this would lead us to
$\dalp \sim 0.00028$ (data), which has to be confronted with the
theory driven estimate $\dalp \sim 0.00016$ (theo)~\cite{KS98,DH98b}.
There is little hope that this precision can ever be reached by an
experiment.\\
\noi
Alternative strategies:\\[-5mm]
\begin{itemize}
\item try to determine $\dahz$ by a direct measurement
\item \underline{simplest:} find the Higgs and determine $m_H$
precisely; this would allow us to determine $\dahz$ from the precise
value of $\sin^2 \Theta^\ell_{\rm eff}$ measured at LEP/SLD.
\item For what concerns theory, which is essentially pQCD,
we need better estimates of uncertainties (scheme dependence, higher
orders etc. which are a problem at $M_\tau$), extended studies in \MOM
schemes~\cite{JT98}, analytic extensions of the \MSb
scheme~\cite{SB98} etc. are needed.
\end{itemize}
\noi
In the meantime I propose to use in phenomenological applications an estimate
of the kind I presented in Sec.~3.

\section*{Acknowledgments}
I thank the organizers of the Symposium RADCOR'98 for the kind invitation and
the excellent hospitality at Barcelona. Furthermore, I thank
S. Eidelman, A. Nyffeler and O. Veretin for stimulating discussions
and A. Nyffeler and O. Veretin for carefully reading the
manuscript.


\section*{References}
\begin{thebibliography}{99}

\bibitem{Rhein} F. Jegerlehner, \NPBPS {\bf 51C}, 131 (1996)

\bibitem{MS98} M. Steinhauser, \PLB {\bf 429}, 158 (1998)

\bibitem{GKL}
S.G. Gorishny, A.L. Kataev and S.A. Larin, \PLB {\bf 259}, 144 (1991);\\
L.R. Surguladze and M.A. Samuel, \PRL {\bf 66}, 560 (1991); ibid. 2416
(1991) (Err);\\
K.G. Chetyrkin, \PLB {\bf 391}, 402, (1997)

\bibitem{EJ95} S. Eidelman and F. Jegerlehner, \ZPC {\bf 67}, 585
(1995); see also:
               H. Burkhardt, B. Pietrzyk, \PLB {\bf 356}, 398 (1995)


\bibitem{ChK95} K.G.~Chetyrkin and J.H.~K\"uhn,
                 \Journal{\PLB}{342}{356}{1995}

\bibitem{Tsai} Y. S. Tsai, \PRD {\bf 4}, 2821 (1976)

\bibitem{EI95} S. Eidelman and V.N. Ivanchenko, \PLB {\bf 257}, 437 (1991);
\NPBPS {\bf 40C}, 131 (1995) and private communication.

\bibitem{aleph} ALEPH Collaboration (R. Barate {\it et al.}),
\ZPC {\bf 76}, 15 (1997); \EPJ {\bf 4}, 409 (1998)

\bibitem{opal} OPAL Collaboration (K. Ackerstaff {\it et al.}),
CERN-EP-98-102, hep-ex/9808019

\bibitem{ADH98} R. Alemany, M. Davier, A. H\"ocker,
                {\it Eur.~Phys.~J.} C {\bf 2}, 123 (1998)

\bibitem{MZ95}  A.D. Martin, D. Zeppenfeld, \PLB {\bf 345}, 558 (1995)

\bibitem{DH98a} M. Davier, A. H\"ocker, \PLB {\bf 419}, 419 (1998)

\bibitem{KS98} J.H. K\"uhn, M. Steinhauser, \PLB {437}, 425 (1998);
see also:\\
K.G. Chetyrkin, A.H. Hoang, J.H. K\"uhn, M. Steinhauser, T. Teubner,
                {\it Eur.~Phys.~J.} C {\bf 2}, 137 (1998)

\bibitem{GKSN98} S. Groote, J.G. K\"orner, N.F. Nasrallah, K. Schilcher,
                 \PLB {\bf 440}, 375 (1998)

\bibitem{DH98b} M. Davier, A. H\"ocker, \PLB {\bf 435}, 427 (1998)

\bibitem{SW96} M.L. Swartz, \PRD {\bf 53}, 5268 (1996)

\bibitem{Erler98} J. Erler, hep-ph/9803453

\bibitem{CM98} A. Czarnecki, W. J. Marciano, BNL-HET-98-43, hep-ph/9810512

\bibitem{MS86} W.J. Marciano, A. Sirlin, \PRL {\bf 56}, 22 (1986)

\bibitem{mqcd3} K.G. Chetyrkin, J.H. K\"uhn, M. Steinhauser,
               \PLB {\bf 371}, 93 (1996);
               \NPB {\bf 482}, 213 (1996); B {\bf 505}, 40 (1997); \\
               K.G.~Chetyrkin, R.~Harlander, J.H.~K\"uhn, M.~Steinhauser,
               \NPB {\bf 503}, 339 (1997)

\bibitem{SVZ} M.A. Shifman, A.I. Vainshtein and V.I. Zakharov,
              \NPB {\bf 147}, 385 (1979)

\bibitem{FJ86} F. Jegerlehner, \ZPC {\bf 32}, 195 (1986)

\bibitem{NPho}
D.J. Broadhurst and S.C. Generalis, Open University preprint OUT-4102-12 (1984)
(unpublished); S.C. Generalis, {\it J. Phys.} {\bf G 15} (1989) L225.

\bibitem{ChGS}
K.G. Chetyrkin, S.G. Gorishny and V.P. Spiridonov, {\it Phys. Lett.} {\bf
160B} (1985) 149.

\bibitem{SCh}
V.P. Spiridonov and K.G. Chetyrkin, {\it Yad. Fiz.} {\bf 47} (1988) 818
[{\it Sov. J. Nucl. Phys.} {\bf 47} (1988) 522]

\bibitem{ST90}
L.R.~Surguladze and F.V.~Tkachov, {\it Nucl. Phys.} {\bf B 331}
(1990) 35.

\bibitem{JT98} F.~Jegerlehner, O.V.~Tarasov,
               DESY 98-093 (hep-ph/9809485).

\bibitem{EJKV98} S.~Eidelman, F.~Jegerlehner, A.L.~Kataev, O.~Veretin,
                 DESY 98-206 (hep-ph/9812xxx).

\bibitem{PDG}
C. Caso et al.(Particle Data Group), {\it Eur. Phys. J.} C {\bf 3},1 (1998)

\bibitem{SB98} S. Brodsky, these Proceedings

\end{thebibliography}
\end{document}